\documentclass[pra,aps,reprint,showpacs,amsmath]{revtex4-1}

\usepackage{bm}
\usepackage{graphicx}
\usepackage{dsfont}

\newcommand{\bra}[1]{\langle #1 | \,}
\newcommand{\ket}[1]{\, | #1 \rangle}
\newcommand{\braket}[2]{\langle #1 | #2 \rangle}
\newcommand{\expv}[1]{\langle #1 \rangle}
\newcommand{\om}{\omega}
\newcommand{\Om}{\Omega}
\newcommand{\ga}{\gamma}
\newcommand{\Ga}{\Gamma}

\newcommand{\De}{\Delta}

\newcommand{\la}{\lambda}

\newcommand{\hlf}{\frac{1}{2}}
\newcommand{\mc}[1]{\mathcal{#1}}
\newcommand{\sig}{\hat{\sigma}}

\newcommand{\hL}{\hat{L}}

\begin{document}

\title{Filtering single atoms from Rydberg blockaded mesoscopic ensembles}

\author{David Petrosyan}
\affiliation{Aarhus Institute of Advanced Studies, Aarhus University,
DK-8000 Aarhus C, Denmark}
\affiliation{Institute of Electronic Structure and Laser, FORTH,
GR-71110 Heraklion, Crete, Greece}

\author{D. D. Bhaktavatsala Rao}
\affiliation{Department of Physics and Astronomy, University of Aarhus,
DK-8000 Aarhus C, Denmark}

\author{Klaus M\o lmer}
\affiliation{Department of Physics and Astronomy, University of Aarhus,
DK-8000 Aarhus C, Denmark}

\date{\today}

\begin{abstract}
We propose an efficient method to filter out single atoms
from trapped ensembles with unknown number of atoms.
The method employs stimulated adiabatic passage to reversibly transfer
a single atom to the Rydberg state which blocks subsequent Rydberg
excitation of all the other atoms within the ensemble. This 
triggers the excitation of Rydberg blockaded atoms to short lived 
intermediate states and their subsequent decay to untrapped states.
Using an auxiliary microwave field to carefully engineer 
the dissipation, we obtain a nearly deterministic single-atom source.
Our method is applicable to small atomic ensembles
in individual microtraps and in lattice arrays.
\end{abstract}

\pacs{32.80.Ee, 
32.80.Qk, 
37.10.Jk, 
37.10.Gh
}

\maketitle

\textit{Introduction.---}
Deterministic preparation of single trapped atoms, or arrays thereof,
is an important task required for many
potentially interesting applications, including quantum computation
and simulation \cite{Jaksch1999,Jaksch2000,Schrader2004,Schneider2012,
OLsimulators,Bloch2008,Weimer2010}. Optical lattice experiments can
realize Mott insulators with single atom per lattice site \cite{RevOL}
with low particle or hole defect probability \cite{Bakr2010,Sherson2010}.
Achieving single-atom occupation of spatially separated microtraps with
high fidelity is more challenging. Several techniques have been explored
to reduce the trap occupation number fluctuations \cite{Mohring2005,%
Gruenzweig2010,Bakr2011,Nikolopoulos2010,Rabl2003,Sherson2012,delCampo2008}

Here we propose a method to filter out single atoms from trapped
ensembles with unknown number of atoms.
Our scheme relies on the strong, long-range interaction between atoms
in high-lying Rydberg states \cite{RydAtoms,rydrev} which blocks resonant
optical excitation of more than one Rydberg atom within a certain interatomic 
distance \cite{Jaksch2000,Lukin2001,Vogt2006,Tong2004,Singer2004,Heidemann2007,%
NatPRLGrangier,NatPRLSaffman,Dudin2012,Schauss2012,%
Robicheaux2005,Stanojevic,Petrosyan,Ates2013,Petrosyan2014}.
We use stimulated adiabatic passage technique (STIRAP) \cite{stirap}
to transfer any one atom of the ensemble from the ground state to
the Rydberg state by a two-photon transition via an intermediate
excited state \cite{Moeller2008,Petrosyan2013,Beterov2011,schempp10}.
This atom then strongly suppresses Rydberg excitation of the other atoms
due to the interaction-induced level shifts. The Rydberg-blockaded atoms
now turn into strongly-driven two-level atoms which rapidly decay
from the (intermediate) excited state back to the initial trapped ground
state \cite{Petrosyan2013} as well as to other untrapped states. After
several excitation and decay cycles, all of the Rydberg-blockaded atoms
are removed from the trap. The reverse adiabatic transfer of the remaining
Rydberg atom back to the initial ground state completes the preparation
of a single trapped atom. We achieve high efficiency and reliability
of the single-atom source by carefully engineering the atomic excitation
and dissipation. The former is controlled by the time-dependent amplitude
and detuning of the laser field, while the latter can be accomplished
by employing an auxiliary microwave field to open appropriate decay
channels to the untrapped states.

\begin{figure}[t]
\includegraphics[width=7.0cm]{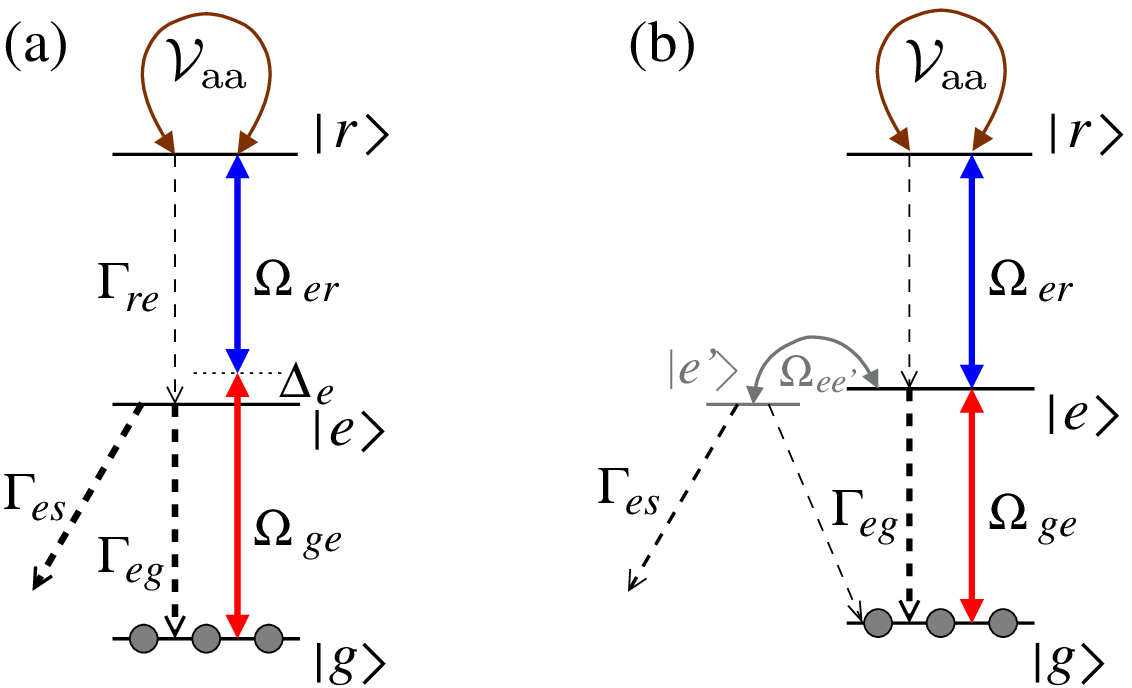}
\caption{
Level scheme of atoms driven by a pair of laser fields with Rabi
frequencies $\Om_{ge}$ and $\Om_{er}$ on the transitions $\ket{g} \to \ket{e}$
and $\ket{e} \to \ket{r}$. The atoms in the ground state $\ket{g}$ are
confined within a volume of linear dimension $L \simeq 2\:\mu$m smaller
than the blockade distance $d_{\mathrm{b}}$ determined by the interaction
$\mc{V}_{\mathrm{aa}}$ between their Rydberg states $\ket{r} \equiv nS_{1/2}$
with large principal quantum number $n$ ($\simeq 50$) and small decay
$\Ga_{re}$ ($\simeq 3\:$KHz) and dephasing $\ga_{r}$ ($\lesssim 0.1\:$MHz) rates.
In (a) the ground state of $^{87}$Rb atoms is
$\ket{g} \equiv 5S_{1/2} \ket{F=1,m_F=1}$ while the intermediate
state $\ket{e} \equiv 5P_{1/2} \ket{F=2,m_F=2}$ decays back to
$\ket{g}$ with the rate $\Ga_{eg} = \hlf \Ga_e$ ($\Ga_e = 36\:$MHz) and
to other untrapped states $\ket{s} \equiv 5S_{1/2} \ket{F=2,m_F=1,2}$ with
the rate $\Ga_{es} = \hlf \Ga_e$ (branching ratio $\Ga_{es}/\Ga_{eg} =1$).
In (b) the transition between the ground
$\ket{g} \equiv 5S_{1/2} \ket{F=2,m_F=2}$ and intermediate
$\ket{e} \equiv 5P_{3/2} \ket{F=3,m_F=3}$ states is closed
($\ket{e}$ decays only to $\ket{g}$ with rate $\Ga_e = 38\:$MHz),
but a microwave field coupling $\ket{e}$ to, e.g.,
$\ket{e'} \equiv 5P_{1/2} \ket{F=2,m_F=2}$ with Rabi frequency $\Om_{ee'}$
can open a weak decay channel $\Ga_{es} = \frac{8 |\Om_{ee'}|^2}{3\Ga_{e'}}
\simeq 2.4 \:$MHz ($\ll \Ga_{eg} = \Ga_{e} + \frac{4 |\Om_{ee'}|^2}{\Ga_{e'}}$)
to the untrapped states $\ket{s} \equiv 5S_{1/2} \ket{F=1,2,m_F=1}$.
The hyperfine and/or Zeeman shifts of states $\ket{s}$ suppress
their coupling to the $\Om_{ge}$ field, while facilitating selective
interaction with a pushing laser of appropriate frequency and polarization.}
\label{fig:alsOmPr}
\end{figure}

\textit{Stimulated adiabatic passage with a single-atom.---}
For what follows, it will be helpful to recall the essence of adiabatic
transfer (STIRAP) of population in an isolated three-level atom using a
pair of coherent laser pulses \cite{stirap}. A laser field with Rabi
frequency $\Om_{ge}$ couples the stable ground state $\ket{g}$ to an
unstable (decaying) excited state $\ket{e}$, which in turn is coupled
to another stable state $\ket{r}$ by the second laser field of Rabi
frequency $\Om_{er}$ [Fig.~\ref{fig:alsOmPr}(a)]. In the frame rotating
with the frequencies of the laser fields, the atom-field interaction
Hamiltonian reads
\[
\mc{V}_{\mathrm{af}} = \hbar \De_e \ket{e}\bra{e} + \hbar(\Om_{ge} \ket{e}\bra{g}
+ \Om_{er} \ket{r}\bra{e} + \mathrm{H. c.}),
\]
where $\De_e$ is the detuning of the intermediate state $\ket{e}$
and we assume two-photon resonance between $\ket{g}$ and $\ket{r}$.
The eigenvalues of $\mc{V}_{\mathrm{af}}/\hbar$ are $\la_0 = 0$ and
$\la_{\pm} = \De_e/2 \pm \sqrt{\Om_{ge}^2 + \Om_{er}^2 +(\De_e/2)^2}$
and the corresponding eigenstates are given by
\begin{eqnarray*}
\ket{\psi_0} &=& (\Om_{er} \ket{g} - \Om_{ge} \ket{r})/\om_0 , \\
\ket{\psi_{\pm}} &=& (\Om_{ge} \ket{g} + \la_{\pm} \ket{e} + \Om_{er} \ket{r})
/\om_{\pm},
\end{eqnarray*}
where $\om_0^2 = \Om_{ge}^2 + \Om_{er}^2$ and $\om_{\pm}^2 = \om_0^2 + \la_{\pm}^2$.
The zero-energy ``dark'' state $\ket{\psi_0}$ does not contain the rapidly
decaying state $\ket{e}$, which, however, contributes to the energy-shifted
``bright'' states $\ket{\psi_{\pm}}$ making them unstable against spontaneous
decay. By adiabatically changing the dark state superposition,
the STIRAP process can completely transfer the population
between the two stable states $\ket{g}$ and $\ket{r}$ without
populating the unstable state $\ket{e}$. To this end, with the atom
initially in state $\ket{g}$, one first applies the $\Om_{er}$ field
($\Om_{er} \gg \Om_{ge}$), resulting in $\braket{g}{\psi_0} = 1$.
This is then followed by switching on the $\Om_{ge}$ field,
leading to $\braket{r}{\psi_0} =  - \Om_{ge}/\om_0$, which results
in $|\braket{r}{\psi_0}| \simeq 1$ when $\Om_{ge} \gg \Om_{er}$.
For sufficiently smooth switching,
$\partial_t \om_{0} \ll \om_0 |\la_{\pm} - \la_0|$,
the atom adiabatically follows the dark state $\ket{\psi_0}$, and the
bright states $\ket{\psi_{\pm}}$, and thereby $\ket{e}$, are never populated.
The reverse process, at the end of which $\Om_{er} \gg \Om_{ge}$,
can transfer the atom in state $\ket{r}$ back to state $\ket{g}$.

\textit{The multiatom system.---}
Consider now an ensemble of $N$ atoms initially in
the ground state $\ket{g}$ confined within a small trap.
All the atoms uniformly interact with two laser fields on the transitions
$\ket{g} \to \ket{e}$ and $\ket{e} \to \ket{r}$ with Rabi frequencies
$\Om_{ge}$ and $\Om_{er}$, as shown in Fig.~\ref{fig:alsOmPr}(a).
For each atom $j$, the interaction Hamiltonian reads
\begin{equation}
\mc{V}^j_{\mathrm{af}} = \hbar \De_e \sig_{ee}^j +
\hbar(\Om_{ge} \sig_{eg}^j + \Om_{er} \sig_{re}^j + \mathrm{H. c.}) ,
\end{equation}
where $\sig_{\mu \nu}^j \equiv \ket{\mu}_{jj}\bra{\nu}$ are the atomic operators.
The intermediate excited state $\ket{e}$, possibly detuned by $\De_e$,
decays with the rate $\Ga_{eg}$ to the trapped ground state $\ket{g}$,
and with the rate $\Ga_{es}$ to another untrapped lower state $\ket{s}$
which is decoupled from the driving lasers.
For completeness, we also take into account the typically much smaller
decay rate $\Ga_{re}$ of the highly excited Rydberg state $\ket{r}$
and its possible dephasing $\ga_{r}$ (with respect to the other states)
due to nonradiative collisions, inhomogeneous trapping potential,
external field noise, laser phase fluctuations or Doppler shifts.
The corresponding Lindblad generators for the decay and dephasing processes are
$\hL_{ge}^j = \sqrt{\Ga_{eg}} \sig_{ge}^j$,
$\hL_{se}^j = \sqrt{\Ga_{es}} \sig_{se}^j$,
$\hL_{er}^j = \sqrt{\Ga_{re}} \sig_{er}^j$, and
$\hL_{r}^j = \sqrt{\ga_{r}} (2 \sig_{rr}^j - \mathds{1}^j)$,
where $\mathds{1}^j \equiv \sig_{gg}^j + \sig_{ss}^j + \sig_{ee}^j + \sig_{rr}^j$.

The long-range atom-atom interactions are described by the Hamiltonian
\begin{equation}
\mc{V}_{\mathrm{aa}}^{ij} = \hbar \Delta(d_{ij}) \sig_{rr}^i \otimes \sig_{rr}^j ,
\end{equation}
where $\Delta(d_{ij}) = C_6/d_{ij}^{6}$ is the van der Waals potential
between pairs of atoms $i,j$ in the Rydberg state $\ket{r}$
separated by distance $d_{ij}$. When the interaction-induced
level shifts $\Delta(d_{ij})$ of states $\ket{r_i r_j}$ exceed
the two-photon excitation linewidth of the Rydberg state $w \simeq
\frac{\Omega_{ge}^2 +\Omega_{er}^2 }{\sqrt{2 \Omega_{ge}^2+\Ga_{e}^2/4}}$
($\Ga_{e} = \Ga_{eg} + \Ga_{es}$), double (multiple) Rydberg excitations
are blocked. We can then define the blockade distance via
$\Delta(d_{\mathrm{b}}) \gtrsim w_{\mathrm{max}}$ leading to
$d_{\mathrm{b}} = \sqrt[6]{C_6/w_{\mathrm{max}}}$. To achieve nearly
complete blockade of more than one Rydberg excitation, we assume
that all the atoms are initially trapped within a volume of linear
dimension $L \lesssim \frac{2}{3} d_{\mathrm{b}}$ corresponding
to $\Delta(d_{ij}) \gtrsim 10 w_{\mathrm{max}} \; \forall \; i,j \in [1,N]$.

\textit{Numerical simulations.---}
We simulate the dissipative dynamics of the system of $N$ atoms
using the quantum stochastic (Monte Carlo) wavefunctions \cite{qjumps},
which allows us to treat nontrivial numbers $N \leq 10$ of strongly-interacting
atoms exactly, without truncating the correspondingly large Hilbert space.
In such simulation, the state of the system $\ket{\Psi}$ evolves
according to the Schr\"odinger equation
$\partial_t \ket{\Psi} = -\frac{i}{\hbar} \tilde{\mc{H}} \ket{\Psi}$
with an effective Hamiltonian
\begin{equation}
\tilde{\mc{H}} = \mc{H} - \frac{i}{2} \hbar \hL^2 ,
\end{equation}
where
\[
\mc{H} = \sum_j \mc{V}_{\mathrm{af}}^j + \sum_{i<j} \mc{V}_{\mathrm{aa}}^{ij}
\]
is the usual (Hermitian) Hamiltonian, while
\begin{eqnarray*}
\hL^2 &=& \sum_j [\hL_{ge}^{j\dag}\hL_{ge}^{j} + \hL_{se}^{j\dag}\hL_{se}^{j}
+ \hL_{er}^{j \dag} \hL_{er}^j + \hL_{r}^{j\dag} \hL_{r}^j ] \\
&=& \sum_j [(\Ga_{eg} + \Ga_{es}) \sig_{ee}^j + \Ga_{re} \sig_{rr}^j +
\ga_{r} \mathds{1}^j]
\end{eqnarray*}
is the non-Hermitian part which does not preserve the norm of $\ket{\Psi}$.
The evolution is interrupted by random quantum jumps
$\ket{\Psi} \to \hL_{\mu \nu}^j \ket{\Psi}$ of individual atoms
with probabilities determined by the corresponding weights
$\bra{\Psi} \hL_{\mu \nu}^{j\dag} \hL_{\mu \nu}^{j} \ket{\Psi}$.
In a single quantum trajectory, the normalized wavefunction
of the system at any time $t$ is given by
$\ket{\bar{\Psi}(t)} = \ket{\Psi(t)}/\sqrt{\braket{\Psi(t)}{\Psi(t)}}$.
The expectation value of any observable $\hat{\mc{O}}$ of the system
is then obtained by averaging over many, $M \gg 1$, independently simulated
trajectories, $\expv{\hat{\mc{O}}} = \mathrm{Tr} [\hat{\rho}\hat{\mc{O}}]
= \frac{1}{M} \sum_m^M \bra{\bar{\Psi}_m} \hat{\mc{O}} \ket{\bar{\Psi}_m}$,
while the density operator is given by
$\hat{\rho}(t) = \frac{1}{M} \sum_m^M \ket{\bar{\Psi}_m(t)}
\bra{\bar{\Psi}_m(t)}$.

Starting from all the atoms in the trapped ground state,
$\ket{\Psi(0)} = \ket{g_1,g_2,\ldots,g_N}$, we assume that during
the evolution the atoms that decay to state $\ket{s}$ decouple
from the laser field and are eventually lost from the trap
(which can be accelerated by a pushing laser).
The probabilities $P_N(n) = \expv{\hat{\Pi}_N^{(n)}}$ for $n$,
out of the initial $N$, atoms surviving in the trap are then given
by the projectors $\hat{\Pi}_N^{(0)} \equiv  \prod_{i=1}^{N} \sig_{ss}^i$,
$\hat{\Pi}_N^{(1)} \equiv \sum_{j=1}^N (\mathds{1}^j - \sig_{ss}^j)
\prod_{i\neq j}^{N} \sig_{ss}^i$, etc. The mean number of trapped atoms is then
$\expv{n} = \sum_n n P_N(n) = \expv{\sum_{j=1}^{N} (\mathds{1}^j - \sig_{ss}^j)}$.
Typically, the initial number of trapped atoms is uncertain, with
the Poisson distribution
\[
P_{\mathrm{init}}(N) = \frac{\bar{N}^N e^{-\bar{N}}}{N!}
\]
around some mean $\bar{N}$. Then, the probabilities $P(n)$ for $n$
atoms surviving in the trap are given by
$P(n) = \sum_N P_{\mathrm{init}}(N)P_N(n)$.

\begin{figure}[t]
\includegraphics[width=8.5cm]{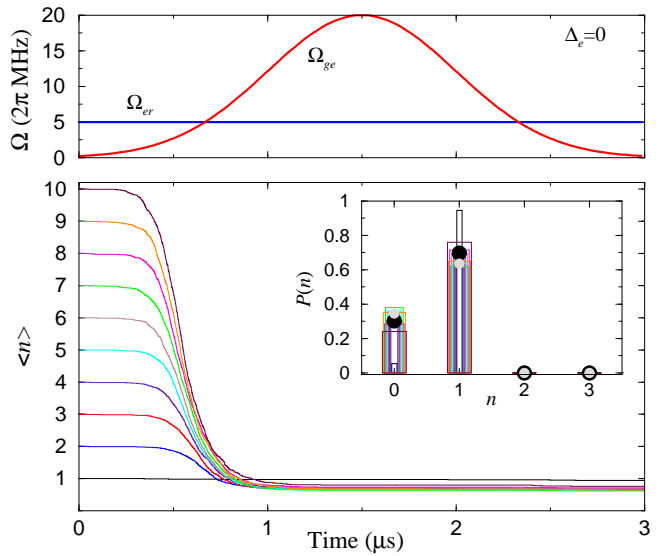}
\caption{
Dynamics of the system, as obtained from $M=200-500$ independent
realizations (trajectories) of the numerical experiment for each
number $N$ of atoms with the level scheme of Fig.~\ref{fig:alsOmPr}(a),
irradiated with two resonant fields, $\De_e=0$, of Rabi frequencies
$\Om_{ge}$ and $\Om_{er}$ shown in the upper panel.
Main panel shows the mean number $\expv{n}$ of surviving atoms in the trap,
starting with $N=1-10$ (from bottom to top) ground state atoms, for $\ga_r=0$.
Inset shows the corresponding probabilities $P_N(n)$ of finding $n$
trapped atoms in state $\ket{g}$ at the end of the process (open bars),
while the black ($\ga_r=0$) and gray ($\ga_r \neq 0$) circles denote 
the $P(n)$ averaged over the initial Poisson number distribution
$P_{\mathrm{init}}(N)$ of the trapped atoms with the mean $\bar{N}=5$.}
\label{fig:Nav}
\end{figure}

\begin{figure}[t]
\includegraphics[width=8.5cm]{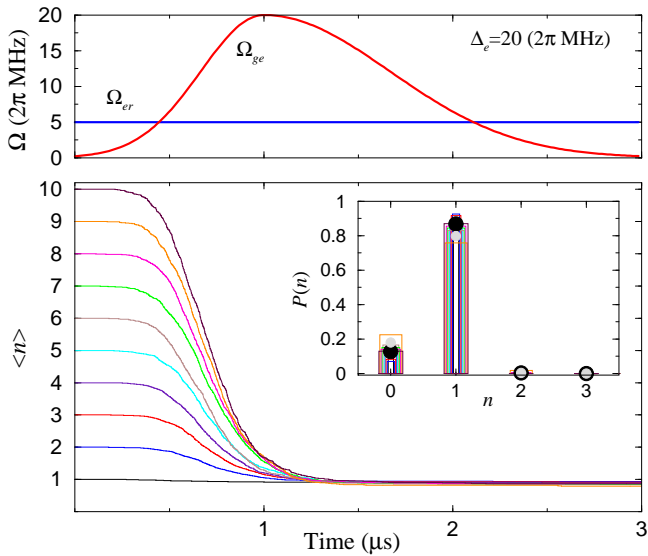}
\caption{
Same as in Fig.~\ref{fig:Nav}, but with finite detuning $\De_e$
and slightly different optimized pulse shape of $\Om_{ge}$.}
\label{fig:NavD}
\end{figure}

\begin{figure}[t]
\includegraphics[width=8.5cm]{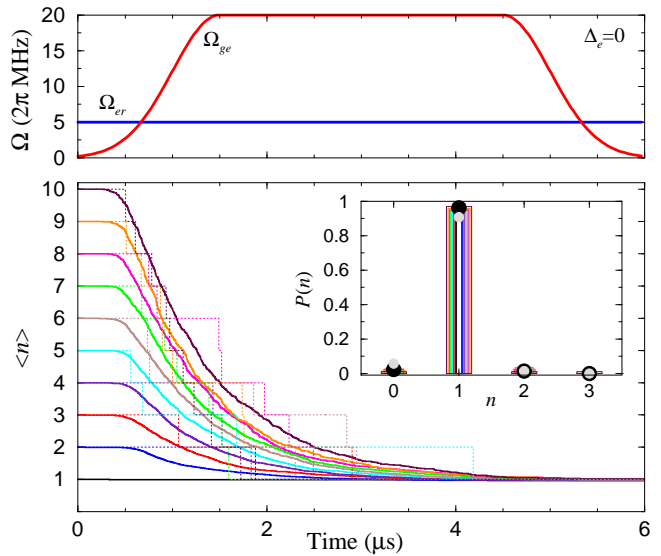}
\caption{
Same as in Figs.~\ref{fig:Nav} and \ref{fig:NavD}, but for the scheme
of Fig.~\ref{fig:alsOmPr}(b) with resonant fields, $\De_e =0$, and
weaker decay to $\ket{s}$ (notice the longer time scale of the process).
Thin dotted lines in the main panel show single trajectories for the
corresponding initial number of atoms $N=1-10$.}
\label{fig:NavG}
\end{figure}

We have performed numerical simulations for various initial
number $N$ of atoms in the trap, results of which are shown
in Figs.~\ref{fig:Nav}-\ref{fig:NavG}.
In all simulations, we first apply the field coupling states
$\ket{e}$ and $\ket{r}$ with the Rabi frequency $\Om_{er}$ which
stays constant throughout the process, and then send a smooth pulse
driving transition $\ket{g} \to \ket{e}$ with the Rabi frequency
$\Om_{ge}(t)$ whose peak value is much larger than $\Om_{er}$. We thus
expect that a single atom in the trap will be adiabatically transferred
from the ground state $\ket{g}$ to the Rydberg state $\ket{r}$ and
block Rydberg excitations of all the other atoms initiating thereby
their optical pumping through the intermediate excited state $\ket{e}$
to the decoupled and untrapped state(s) $\ket{s}$. At the end of the
process, when $\Om_{ge}$ is reduced to zero, the remaining Rydberg-state
atom will be transferred back to the trapped ground state $\ket{g}$.

\textit{Results.---}
Consider first the resonant excitation, $\De_e = 0$, of the atoms
with the $\Om_{ge}$ and $\Om_{er}$ fields shown in the upper panel
of Fig.~\ref{fig:Nav}. The branching ratio for the decay of the atoms
from the intermediate excited state $\ket{e}$ to the untrapped state
$\ket{s}$ and back to $\ket{g}$ is large $\Ga_{es}/\Ga_{eg}=1$,
as per Fig.~\ref{fig:alsOmPr}(a), while we neglect for the moment
the dephasing of the Rydberg state, $\ga_r=0$. We find that even before
the $\Om_{ge}$ pulse attains its maximum, most of the atoms are optically
pumped into the untrapped state(s) $\ket{s}$, while the mean number
of atoms in the trap drops below unity: $P(0) \simeq 0.3$
and $P(1) \simeq 0.7$, while $P(n>1) \simeq 0$. The reason for
this unexpected and unfortunate behavior is that each quantum jump
of a blocked atom from state $\ket{e}$ to either $\ket{g}$ or $\ket{s}$
projects a single atoms to the Rydberg state $\ket{r}$ \cite{Petrosyan2013}
whose overlap with the dark state
$|\braket{\psi_0}{r}|^2 = \Om_{ge}^2/(\Om_{ge}^2 + \Om_{er}^2)$
is smaller than unity, unless $\Om_{ge} \gg \Om_{er}$. Since
for $\Om_{ge} \lesssim \Om_{er}$ the Rydberg atom is not in the
dark eigenstate, but has large overlap with the bright eigenstates
as well, $|\braket{\psi_{\pm}}{r}|^2 = \Om_{er}^2/(2\Om_{ge}^2 + 2\Om_{er}^2)$,
it is coupled to the state $\ket{e}$ from where it can decay to
$\ket{g}$ or $\ket{s}$. In a closed three-level system \cite{Petrosyan2013},
the decay to $\ket{g}$ would simply restart the excitation process of this
or any other atom by the fields $\Om_{ge}$ and $\Om_{er}$. In the present case,
however, the decay of all the atoms but one to the decoupled state
$\ket{s}$ may leave the last atom in the Rydberg state $\ket{r}$ before
the $\Om_{ge}(t)$ pulse reaches its peak value $\Om_{ge} \gg \Om_{er}$.
The probability of loosing this last remaining atom can thus be estimated as
\[
\frac{\Ga_{es}}{\Ga_{es} + \Ga_{eg}} (1 - |\braket{\psi_0}{r}|^2)
\simeq \frac{\Ga_{es}\Om_{er}^2}{(\Ga_{es} + \Ga_{eg})(\Om_{ge}^2 + \Om_{er}^2)},
\]
which, with $\Ga_{es} = \Ga_{eg}$ and $\Om_{ge} \simeq \Om_{er}$ is about
$1/4$, consistent with $\expv{n}$ and $P(0)$ in Fig.~\ref{fig:Nav}.

Including now in the simulations realistic dephasing, $\ga_r \neq 0$,
amounts to additional coherence relaxation between the long-lived states
$\ket{g}$ and $\ket{r}$ with the rate $\ga_{rg} = \hlf \Ga_{re} + 2 \ga_r$.
As is well known \cite{stirap}, the STIRAP in a three-level atom is
very sensitive to such dephasing, since the decoherence of the dark-state
superposition leads to additional population of the bright states
\cite{Petrosyan2013}, which in the present system can decay
to the untrapped state(s) $\ket{s}$. Hence, in the inset of
Fig.~\ref{fig:Nav} we observe further decrease of the survival
probability $P(1) \simeq 0.63$ for a single trapped atom.

To increase the single atom survival probability $P(1)$, we should
therefore postpone the decay of atoms from state $\ket{e}$ to $\ket{s}$
until $\Om_{ge}$ is much larger than $\Om_{er}$.
We thus take a finite detuning $\De_e$ of intermediate level $\ket{e}$
and switch-on $\Om_{ge}$ slightly faster. The results of simulations are
shown in Fig.~\ref{fig:NavD}, were we indeed notice significant
improvement of the single-atom filtering, $P(1) \simeq 0.87$ 
or $0.8$ with additional dephasing $\ga_r$.
These results are obtained after some optimization of both $\Om_{ge}$
and $\De_e$, but for the given atomic level scheme further substantial
improvements do not seem possible.

To achieve nearly perfect filtering, we now consider the closed three-level
system of Fig.~\ref{fig:alsOmPr}(b), where state $\ket{e}$ decays spontaneously
only to the trapped state $\ket{g}$, while we can engineer an additional decay
channel to the untrapped states $\ket{s}$. To this end, we use a microwave
field which weakly couples $\ket{e}$ to an auxiliary state $\ket{e'}$
with the Rabi frequency $\Om_{ee'}$ that is small compared to the total
decay rate $\Ga_{e'}$ of $\ket{e'}$. Upon adiabatic elimination of $\ket{e'}$,
we obtain a small additional decay on the transition $\ket{e} \to \ket{g}$
with rate $4 |\Om_{ee'}|^2/3\Ga_{e'}$, and a decay to the untrapped states
with rate $\Ga_{es} = 8 |\Om_{ee'}|^2/3\Ga_{e'}$ which can be
controlled via the microwave field intensity $|\Om_{ee'}|^2$
(the numerical prefactors are related to the branching ratio
of the decay of state $\ket{e'}$ to $\ket{g}$ and $\ket{s}$,
$\Ga_{e'g}/\Ga_{e's} = 1/2$). Figure~\ref{fig:NavG} shows the
results of our simulations demonstrating dramatically improved single
atom filtering, $P(1) \simeq 0.96$ [$P(0) \simeq 0.02$ and
$P(2) \simeq 0.01$, $P(3) \simeq 2\times 10^{-4}$],
at the expense of a longer duration of the process.
Even sizable dephasing $\ga_r$ of the Rydberg state with respect
to the ground state does not significantly reduce the efficiency
of the single-atom filtering, $P(1) \simeq 0.92$ [$P(0) \simeq 0.05$],
which is due to the robustness of adiabatic passage in a Rydberg-blockaded
ensemble composed of atoms with nearly-closed transitions \cite{Petrosyan2013}.

We finally note that in obtaining the averaged probabilities $P(n)$
of the final number of atoms, we assumed the initial Poisson distribution
$P_{\mathrm{init}}(N)$ of the atom number in the trap with the mean
$\bar{N} =5$, for which $P_{\mathrm{init}}(0) \simeq 0.007$ and
$\sum_{N>10} P_{\mathrm{init}}(N) \simeq 0.02$, and performed simulations
for $N \leq 10$. The initial 2\% population of the trap not included
in our simulation can only increase the final single atom probability.
Our method should work even better for larger initial number of atoms in
the trap, which is, however, prohibitively difficult to simulate numerically.

\textit{Summary.---}
To conclude, we have proposed and analyzed a viable scheme to filter
out single atoms from mesoscopic ensembles containing unknown number
of initially trapped atoms. We have shown that the combination of
two-photon adiabatic transfer of the atoms to the strongly interacting
Rydberg state and carefully engineered decay of atoms from the
intermediate excited state can lead to nearly-deterministic production
of single trapped atoms in individual microtraps or optical lattice
potentials, which is an important prerequisite for quantum computation
with atoms serving as qubits or quantum simulations of many-body
spin systems. Moreover, our studies represent an important and interesting
new facet of the very active field of dissipative generation of states
and processes in open many-body systems \cite{Diehl2008,Verstraete2009}.


\end{document}